\documentclass[12pt]{article}
\usepackage{mathtext}
\usepackage{graphicx}
\usepackage[english]{babel}
\usepackage[T2A]{fontenc}
\usepackage[utf8]{inputenc}
\usepackage{amsfonts}
\usepackage{amssymb}
\usepackage{amsmath}
\newcommand{\ar}{\longrightarrow}

\newcommand{\la}{\lambda}

\begin{document}
\title{How to check quantum mechanics independently (Reply to arXiv:1505.04293)}
\author{Y.Ozhigov\thanks{MSU, VMK, ozhigov@cs.msu.su}}
\maketitle

\begin{abstract}
This is the reply to the paper of Andrei Khrennikov arXiv:1505.04293 in which he expresses dissatisfaction with that the rough data in quantum experiments is not easily available and compares it with the open rough data in genetics. 
I try to explain why quantum experiments rough data is closed and why it differs radically from the biological. I also tried to answer the more thorny issue: is it possible to check quantum mechanics independently of other humans, e.g. trusting nobody.
 \end{abstract}

\section{Why do we need to check quantum mechanics at all}

I read a recent article by Andrei Khrennikov \cite{1505.04293} dedicated to the acute question concerning the precise test of quantum theory. No one doubts the greatness of this theory, but the check is always necessary and this article prompted me to answer the question raised on the basis of my knowledge.

Quantum mechanics is not just the domain of science, but the modern scientific world view related to Nature in general. It aspires to serve as a universal language on which humans can describe any phenomena, including biological; such a global project as quantum computer eloquently testifies this. The power of the quantum description of the world and its beauty will give us reason to hope that this world view will really help to overcome monstrous differentiation of scientific knowledge. 
 
 The accuracy of quantum predictions amazes any imagination, for example, the calculation of the constant of electromagnetic interaction is obtained with the same error, as if to measure the distance from New York to Chicago to within the thickness of a human hair - (Feynman, an approximate translation, see \cite{Fe}). 

By the accuracy we mean, of course, in principle possible accuracy. It is limited by the uncertainty relation and the context in which experiments are conducted. Quantum theory, as the only first-principles theory, is based on the test constants. These constants are not calculated. They can be only selected on the basis of typical experiments.

 With several key tests, it costs nothing to find numerical values, which makes the exact match for any absurd ``theory'', but only for this narrow class of experiments. The power of quantum mechanics is that if we find these constants to a few first-principle experiments, the derived values give the coincidence with predictions for hundreds of other experiments with mad accuracy. There is no other such theory.

That is why it is important to check all the quantum experiments, and the question posed by Khennikov in his work, is perfectly legal.

There is also more radical formulation: if quantum mechanics pretends to be the queen of sciences, how can we make sure that it is true? Logically, we can not even prove that it will be never found the correct mathematical proof of the equality $ 2 \times 2 = 5 $ (Gödel's theorem about the proof of consistency). Therefore, only an experiment can be the criterion. 
But we must remember that nothing is free, and for the power of quantum theory we pay by the difficulty of getting its predictions. Nothing than cannot be physically computed can be compared with experiment. A. Khrennikov in the work \cite{1505.04293} talks mainly about photons, and I try to explain difficulties of direct examination of quantum electrodynamics that is the most profound part of quantum theory. I urge the reader to gain some patience: the known (but rarely repeated) things should be recollected, because it is important for us now.

At first we need a classical state of electromagnetic field, e.g. a function of the form 
\begin{equation}
(\bar A,\phi ): R^3\ar R^4=R^3\times R,
\label{class}
\end{equation}
which associates with each spatial point the pair $(A,\phi )$ of vector and scalar potential of the field in this point. Already here we meet the difficulty, since all the space $R^3$ is inaccessible for us, we can work in only its finite area. Further, the practical storage of this function (\ref{class}) requires that we divide (limited area) of the state $R^3$ to the finite set of cubes and get thus the discrete representation with the grain of spatial resolution $\delta x$ (for the dynamics we also need the grain of time resolution $\delta t$). The choice of the grain determines probe values of charges and masses, which can change in renormalization though the real magnitudes founds by these values can only become more accurate. 

This is not all. We have to make discrete not only coordinate space but also the set of values of the field: the vector $(A,\phi )$ must take values from the finite set. This vector plays the role of the coordinates of imaginary two oscillators (corresponding to basic polarizations)  connected with the field in each point, hence we must make it discrete as well. 

Suppose that we have overcome these difficulties (although it is really made only for isolated particles, and that with some approximation, for which a system of programming Maple was created) and we already have a discrete representation of the classical spaces. We thus obtain (potentially) the set ${\cal F}$ of all classical states of the field. The power of this set is monstrous, and is equal to $K^L$, where $L$ is the number of points in the discrete representation of coordinate space $R^3$, $K$ of the field oscillators.  

How to apply quantum mechanics to the field? It is needed to quantize the field, e.g. to consider linear combinations of its classical states with the different amplitudes. There are functions of the form
\begin{equation}
\Psi :\ {\cal F}\ar C
\label{quant}
\end{equation}

The object (\ref{quant}) can only belong to our imagination, and even then not completely. This could put an end, if it was not about quantum mechanics. It copes with the quantum state of the field with its inherent ease, introducing the concept of the photon. It looks like this. At first we introduce the transformation of the form ${\cal F}\ar{\cal F}$, called  the ``passage to the impulse basis'', which is the Fourier transform. From the view point of the space of quantum states this ``passage to the impulse basis'' is the simple permutation of basic vectors, e.g. the particular case of unitary operator. 

If oscillators $A(R),\phi (R)$ for the different $R$ depend from each other even if there are no charges, oscillators of the impulse representation will be independent in the free field. It means that we can muliply the wave functions. A photon with the impulse $p$ and the polarization $\epsilon$ is such a state of the field, in which the oscillator of the impulse representation, corresponding to $p$ and $\epsilon$, is in the first excited state whereas all others are in the ground state.  We denote the pair $A,\phi$ by $a$ and for simplicity assume that the polarization and impulse are contained in the vector $k$. The ground state (vacuum) of the field is such a state, in which all impulse oscillators are in the ground states, that is its wave function in coordinate form has the Gaussian form (I omit constants for simplicity):
\begin{equation}
\begin{array}{ll}
|\Psi_{vac}(a)\rangle &= \prod_k exp(-\int a(R_1)e^{-ikR_1}dR_1\int a^*(R_2)e^{-ikR_2}dR_2)=\\
&exp\sum_k(-\int a(R_1)a^*(R_2)e^{-ik(R_1-R_2)}dR_1dR_2=\\
&exp(-\int a(R)a^*(R)dR).
\end{array}
\label{vak}
\end{equation}

Here in the product and integrals arguments belong to the discrete set of values; we also use the great principle of interference, according to which we ignore all finite numbers when comparing them with the infinite. If for some classical state $a$ of the filed its classical amplitude $a(R)$ noticeably differs from zero on the set of the nonzero measure, its quantum amplitude will be negligible thanks to fast decreasing of Gaussian that corresponds to the intuitive representation of the vacuum. 

If there is one photon in the field, we must multiply (\ref{vak}) to $a_{k_0}$, where $k_0$ is its impulse and polarization together. That is for one photon the amplitude corresponding to the classical state of the field $a(R)$ equals 
\begin{equation}
|\Psi (a)\rangle=\int a(R)\ e^{-ik_0R}dR\ exp(-\int |a(R)|^2\ dR).
\label{one_photon}
\end{equation}
This amplitude will not be negligible only for such states of the field: 
\begin{equation}
a_{1\ photon}(R)=\epsilon\ e^{ik_0R}
\label{class_photon}
\end{equation}
that is often called the ``wave function of photon'', though its real wave function has the form  (\ref{one_photon}). 

Let you try to do the same for two photons, if you fail read the book \cite{AB}. 

Now we see what the wave function is and how to work with it.  We need to take a number of conventions that have no correct justification in terms of mathematics, mostly on ignoring negligible summands. This is not a ``carelessness '' in the calculation. This is the natural assumption that is necessary to obtain the result from calculations.

Why these assumptions are necessary? Look at the object  (\ref{quant}) once again. To store it in the memory of imaginary computer this memory must have the size 
\begin{equation}
G=N^{K^L}
\label{super_exp}
\end{equation}
where (super-super-googol) $G$ is the number of all possible ``wave functions'' of the field.  Of course, this facility can not accommodate not just any whichever computer, but the entire universe. This means that no statistical sample can be representative without additional extremely hard assumptions about a set of statistics. Moreover, these assumptions can not even be articulated as the tough conditions, such as upper and lower bounds on the sensitivity, etc.
I have not yet touched here the role of the individual in quantum experiments. The fact that the representation of the field as a superposition of Fock states of the form
$$
|\Psi\rangle = \sum\limits_{n\in\{ classical\ n-photon\ states\} }\la_n|n\rangle
$$
depends on who does it. If I'm standing on a mountain top, I have my own representation of the field. But if you fly in an airplane with a speed of 900 kilometers per hour near me, you will have slightly different representation of the same object, because for you it will be a little different field due to relativistic effects. Statistics photodetector clicks can vary depending on the observer. I still ignore the fact that the space is nonuniform and is a 4-dimensional manifold (I apologize to experts in string theory) so that the Lorentz transformation must be carried out in each local card and not in the whole space (where imaginary ``paradoxes'' of the general theory of relaticity comes from, like the fall of the cross, and others.), and it may be relevant in experiments with photons. That's what ``account of space and time'' means in theory.

The same is true for the experiments. There is no ``pure'' experiment, regardless of the theory - this is an illusion. The experiment can be set only if you agree with the language in which the theory is formulated. The language of quantum mechanics rests on wave functions - we see what it is for the field.

Ab initio predictions of quantum theory can exist only for very simple systems, for example, a single photon, and even with the rigid fixation of the grain of rezolution and the associated accuracy. If we make the grain smaller, we can distinguish between the photons, which would merge in the case of bigger grain (rough experiment). As a result, interference effects, resulting from identity of photons have been observed on the rough device  disappear (though on the rough device the decoherence will be bigger that destroys the interference).

For more complex processes are always used ersatz - presentation of photons, which is associated with the imposition of new restrictions on the collection of statistics. If the experimenter publish `` rough results of the experiments '', before the necessary cleaning, it will be immediately accused of a lack of professionalism, because on the basis of rough results nobody can distinguish a conscientious experimenter from a rogue.

A. Khrennikov compares quantum experiments with biological, where ``decrypted'' genomes are posted on the website (\cite{pdb}). This can not serve as an argument. Firstly, none of the genomes is ``decrypted'' - they are sequenced; decryption - in the sense in which things are called in quantum mechanics, is the ability to build a living creature by artificial means, that is too far from us. Secondly, the number of possible genomes even cannot be compared with the volume of the set of quantum states, which is sampled for rough experimental results. In the case of genomes we can make statistically reliable conclusions (not always),  we can not assert the same for rough results in quantum physics, where the statistical reliability can only be achieved as a result of not just the super hard selection of results, but some a priori assumptions about them.

I described the small part of the real situation with the predictions of quantum mechanics, with which I am slightly familiar (others may say more). Why Quantum Mechanics has such authority? The fact is that if you accept quantum methodology, even if you do not take into account all the factors, the result will be still satisfactory. And if there is a discrepancy with the experiment, you should try to take into account the factors that you have ignored, this always helps. Experience shows that after the careful examination of any process we can make a quantum calculation, and the results will coincide with the experimental data highly. The process under consideration should be formulated very precisely that it is not always easy, and sometimes just as hard to get as the ab initio predictions of quantum theory.	

\section{Individuality of quantum experiments}

Let's go back to the lack of ``rough results of quantum measurements''. It really is. I have already partly explained why this is the case. However, the problem is more serious. This is - the problem of trust. Imagine that you get ``rough results of quantum experiments'', for example, a photodetector clicks. How can you make sure that there is no cheating? For statistics of biphotons you can do so. Alice generates the same pure quantum states: either only non entangled or only entangled,  (it is convenient to work with the polarization, because there is no above-mentioned problems with the space-time part, only you need is to choose the appropriate time and space windows), and sends the results of the arbitral judge. Bob, applying quantum tomography (\cite{tom}), determines what state generates Alice and also sends the output to arbitrator.

This only works for simple quantum states, for which there is no problem of searching all possible classical states of the detector. But the super-exponential growth in complexity (\ref{super_exp}) sets such a low limit of the size for such testing protocols that cases even slightly more significant become unavailable.

Can you check that the experimenter tells you, regardless of him (her)? You will have to reproduce the experiment itself, including all the equipment, its configuration, reproduce all the experimental conditions, including the weather outside. In the present level of technology it is unthinkable if you only are not involved in this as a professional. You will have to trust the experimenter, and be satisfied with only the remote verification of output of such experiments. The direct check is possible only for quantum devices with high intensity, such as many photon lasers, which you can try yourself, personally convinced of their characteristics, such as the degree of concentration of the beam.

Even for a little more subtle devices, such as single-photon lasers, it's hard to do it yourself; try to check is it the single-photon laser or not (strictly speaking, these is no single photon laser). This seemingly inevitable dependence on the opinions of others did not inspire any optimism. In complex  cases, you will always have to rely on someone words. But is it so?

There is a unique still imaginary device, perfect in the logical sense that, which if we only could create it, makes us totally independent in the verification of quantum mechanics. This is the Feynman Quantum Computer. I do not need the assistance of any experimenters to make sure that he is on my desk. I even do not need that it stands on my table, it can stand for 1,000 km far from me, I can test it remotely. Everything is very simple. I generate a circuit of functional elements (this is a simple code of ones and zeros) of length 500. This code encodes the boolean function $f$ of 60 arguments. I do it by hand because  I also can not quite trust in my laptop. And send the code to the object, which represents himself a quantum computer. If it can solve the equation $f(x)=1$ in the available time (I have chosen the number 60 so that the time of classical bruit force through $2^{60}$ numbers with the clock rate of modern computers is unavailable even with parallel processing, whereas for $2^{30}$ it is available). If I get a solution of the equation $ f (x) = 1 $ in the foreseeable future (in 3 days, during that time I'll go to the sea to swim a little), I will be able in one day to manually determine whether this is right or not. That's the whole verification. Cheating is excluded. Grover's algorithm, which Feynman quantum computer must be able to fulfill requires the root of classical time to find the solution. A cheater should be able to solve the problem of $ P = NP (?) $ in the positive sense. If someone loves a factorization of integers, you can use this option: time difference will be almost exponential, but also a barrier to fraud is likely to be lower.

You can know nothing about quantum mechanics. You just need to be able to carry out 500 operations  without error (for verification). And you can test quantum mechanics so that best experts  will not be able to fool you, even if they set this goal.

Quantum computer is a real project that pretends to test the most significant of the theories created by humans.  Feynman version of it repeats the logic of microelectronics, and it is now commonly meant by the name ``quantum computer''.  For it the most of the purely mathematical problems are solved; though in experiments we learn a lot about the main enemy of quantum states - decoherence. Thus, a quantum computer of Feynman type radically solves the problem of verification. Don't you think it's too good to be true?

I adhere to the extreme point of view: Feynman version of quantum computer can not be scalable. Physically realizable can be only those processes, which can be modelled by effective classical algorithms. Therefore, the presence of fast quantum algorithms means that decoherence will inevitably destroy any attempt to make the device, which outperforms a classical computer. All experiments were known up to this point, speak in favor of this. This view point can be called physical constructivism; and there is only one way to disprove it - by building a scalable quantum computer. Try to do it.

There are attempts to formulate the concept of a quantum computer of not Feynman type, such as a biological quantum computer. "The quantum computer works in the head of each of us, we just have to figure out how" - these words of Kamil Valiev, perhaps point at a new, intriguing turn in the rapid development of quantum mechanics that we are experiencing. Biological applications of quantum theory can help us to verify this great thing independently that automatically would remove the sharpness of question that we are trying to answer.

\section{Instead of conclusion}

Andrei Khrennikov in \cite{1505.04293} raised a fundamental problem, and I tried to answer the question of rough results of quantum experiments, and the more delicate question, he has not formulated, but which is appropriate - an independent verification of the quantum theory. Unfortunately, my answer is unlikely to satisfy a mathematician, but what to do - that is the reality.

For the luxury of knowledge about how the world works, we have to pay the price, and this price - a lack of confidence in the correctness of our conclusions. As a consolation, one can only remember that this confidence is not possible within the framework of pure mathematics. Trying to verify the accuracy of quantum physics without the trust between us, looks like the desire of a kitten to catch  of its own tail. We do not get rid of the need to trust others, and the difficulty is only to optimize the possible consequences of this trust.

\end{document}